# Substrate-mediated Borophane Polymorphs through Hydrogenation of Two-dimensional Boron Sheets

*Yuchong Kang[1], Xiaoyun Ma[1], Jing Fu[1], Kun Yang[1], Zongguo Wang[3], Haibo Li[1], Wei Ma[1]\* and Jin Zhang[2]\**

[1] Ningxia Key Laboratory of Photovoltaic Materials, School of Materials and New Energy, Ningxia University, Yinchuan, Ningxia 750021, P.R. China

[2] Max Planck Institute for the Structure and Dynamics of Matter and Center for Free-Electron Laser Science, Luruper Chaussee 149, 22761 Hamburg, Germany

[3] Computer Network Information Center, Chinese Academy of Science, Beijing 100190, China

\* Corresponds to mawei@nxu.edu.cn; jin.zhang@mpsd.mpg.de

ABSTRACT:

Two-dimensional boron monolayer (borophene) stands out from the two-dimensional atomic layered materials due to its structural flexibility, tunable electronic and mechanical properties from a large number of allotropic materials. The stability of pristine borophene polymorphs could possibly be improved via hydrogenation with atomic hydrogen (referred to as borophane). However, the precise adsorption structures and the underlying mechanism are still elusive. Employing first-principles calculations, we demonstrate the optimal configurations of freestanding borophanes and the ones grown on metallic substrates. For freestanding $\beta_{12}$ and $\chi_3$ borophenes, the energetically favored hydrogen adsorption sites are on the top of the boron atoms with CN=4 (CN: coordination number), while the best absorption sites for α' borophene are on the top of the boron atoms with CN=6. With various metal substrates, the hydrogenation configurations of borophene are modulated significantly, attributed to the chemical hybridization strength between B $p_z$ and H s orbitals. These findings provide a deep insight into the hydrogenating borophenes and facilitate the stabilization of two-dimensional boron polymorphs by engineering hydrogen adsorption sites and concentrations.

**KEYWORDS: Borophene, Hydrogenation configurations, Substrate modulation, Two-dimensional materials, Density functional theory**

Borophene has been attracting substantial interest in the field of two-dimensional (2D) materials because of its enormous number of allotropes[1–3] and abundant intriguing properties[4,5]. The appealing electronics and mechanical characters of borophenes are tunable for different borophene polymorphisms[6,7]. For instance, the structural fluxionality of borophenes favors structural phase transitions under tension, resulting in small breaking strains yet highly ductile breaking behavior[8]. Previously, Penev *et al.* investigated the polymorphism of 2D boron sheets, contributing to a thorough exploration of the configurational space[9]. From the practical point of view, the versatile properties of borophenes are very profitable for energy storage materials[10–12], catalysis[13], battery[14,15], and biomedical applications[16,17]. Due to the intrinsic phonon-mediated superconductivity induced by strong electron-phonon coupling, 2D boron sheets could be utilized as superconductor with critical temperature of $T_c \approx 10-20$ K[18]. Liu *et. al.* theoretically investigated the possible fabrication methods of 2D boron sheets on metal substrates for the first time, proposing feasible approaches to synthesize 2D boron sheets[19]. Recently, a number of phases of the two-dimensional boron monolayers were successfully fabricated and characterized on metal substrates (e.g., Ag, Au, and Cu) by several independent groups[20–24]. Interestingly, the energetically preferred polymorphs on substrates are found to be sensitively dependent on metal substrates[25].

Notably, the pristine borophenes are susceptible and easy be oxidized in air, hampering both characterization and practical use in electronic devices[26-28]. To improve the stability, chemical passivation provides an effective method to suppress ambient oxidation[29,30], facilitating the separation of borophenes from metallic substrates. Hydrogenated graphene has been explored based on the hydrogen surface concentration and hydrogen passivation in graphene leads to a tunable bandgap[31]. Very recently, Li *et al.* successfully synthesized borophene polymorphs sheets by hydrogenating borophenes with atomic hydrogen experimentally[32], and proved low oxidation rates after ambient exposure, pointing a promising route to low-dimensional boron sheets[33,34]. Thereafter, Xu and coauthor investigated a variety of borophane polymorphs on Ag substrates by first-principles calculations and revealed that the stability of borophane polymorphs depends on hydrogen pressure[35]. However, it is still elusive about the role of hydrogenation in enhancing the stability of boron sheets and the microscopic mechanism of substrate-modified borophane configurations.

In this work, we unravel the favored hydrogen adsorption sites of different borophene allotropes (e.g., $\beta_{12}$, $\chi_3$, and $\alpha'$) with various element ratios of boron and hydrogen atoms. Most importantly, our results establish a concrete link between the optimal hydrogen adsorption sites and the chemical hybridization of borophane, providing paramount insights for improving stability and reducing the oxidation rates of borophenes. Our further results demonstrate that the metal substrates are capable of modifying the optimal hydrogen adsorption sites significantly.

The configurations of borophene can be viewed as a triangular lattice with varying concentrations of hollow hexagons[36], and different hollow hexagons of borophenes are strongly correlated with coordination number (CN)[37] of boron atoms. To obtain the energetically favorable hydrogen adsorption sites of borophene, we focus on several experimentally realized phases of 2D boron sheets: $β_{12}$, $χ_3$, and α' etc. Among them, freestanding $β_{12}$ and $χ_3$ borophenes are planar, while α' borophene hosts an out-of-plane buckled configuration. For clarity, we name the various absorption sites as $H_x$ and $H_{xy}$, where x indicates the sites where hydrogen atoms are adsorbed on the top of boron atoms with CN=x and $H_{xy}$ means the hydrogen atoms locates on the top on the bridge sites between the boron atoms with CN=x and y. For instance, $H_{45}$ represents that the hydrogen atom is adsorbed on the top of the bridge sites of two boron atoms with CN=4 and 5. To obtain more accurate results, both the atomic positions and lattice constants are fully relaxed for the optimal hydrogen adsorption sites and concentrations.

The most energetically favorable hydrogen absorption sites of freestanding $β_{12}$, $χ_3$, and α' borophanes with different element ratios of hydrogen and boron have been provided by comparing the adsorption energies of all possible adsorption configurations (Figure S1-S6, S8-S10), which is defined as

$$E_{\text{ads}} = \left(E_{\text{total}} - E_{\text{primitive}} - n \times E_{\text{H}}\right)/n \qquad (1)$$

where $E_{\text{total}}$, $E_{\text{primitive}}$, and $E_{\text{H}}$ denote the total energies of borophanes, pristine borophenes, and adsorbed hydrogen atoms, respectively; and n is the number of hydrogen atoms.

For medium element ratios (0.2, 0.25, and 0.125 for $β_{12}$, $χ_3$, and α' borophenes, respectively), associated with configurations of one hydrogen atom adsorbing on the unit cell of borophenes, the optimal adsorption sites of the freestanding $β_{12}$, $χ_3$, and α' borophenes are $H_4$, $H_4$, and $H_6$, with $E_{\text{ad}}$ of -3.60 eV, -3.30 eV and -3.15 eV respectively. For the borophene polymorphs, the hydrogen adsorption leads to buckled structures, as exhibited in Figure 1. In the case of $\beta_{12}$ and $\chi_3$ borophenes, the optimal hydrogen adsorption sites are at the top of the boron atoms with the smallest CN. In contrast, for α' borophene, hydrogen atoms are favored to be adsorbed on the boron atoms with the largest CN, yielding $H_6$ sites as the most stable configuration (panel c).

Figure 1d-f depict the band dispersions of the three energetically stable borophanes with medium element ratios. Obviously, there are several bands crossing the Fermi level, suggesting that the metallic band features of $\beta_{12}$, $\chi_3$, and α' borophanes are preserved after hydrogenation. Compared with the freestanding $\beta_{12}$ sheet, the Dirac cone in the Γ-X direction disappears when adsorbing hydrogen atoms. It might originate from the intensity of chemical hybridization between boron sheets and hydrogens. In addition, the Dirac points could be varied after being placed on metal substrates to compensate for the electron deficiency of borophanes.

We next study the underlying mechanism of these borophane structures by estimating the charge transfer between boron and hydrogen atoms through Bader charge analysis[21,38] (Table S1), which implies the strength of Coulomb interaction and formation of the B-H bonds. In this regard, the number of transferred electrons, which determines the strength of B-H bonds, are expected to be a quantity that dominates the polymorph-dependent adsorption sites. The largest transferred electrons are found when H atoms are adsorbed on $H_4$ sites (0.39 and 0.43 *e* for $\beta_{12}$ and $\chi_3$ borophanes, respectively). As for α' borophene, the $H_5$ configuration with higher absorption energy is accompanied with the largest amount of charge transfer. Hence, there is no obvious dependence of the optimal borophanes on the value of transferred electrons.

From Figure 2, we find that not the electron transfer but rather the intensity of chemical hybridization[39] between boron sheets and hydrogen atoms determines the choice of the stable hydrogen adsorption sites. For $\beta_{12}$ borophene, it is the out-of-plane delocalized $p_z$ orbitals of boron atoms that participate considerably in hybridization with s orbitals of hydrogens, as evidenced by resonance peaks in the vicinity of the Fermi level[25]. Especially, for the $H_4$ structure, the B $p_z$ orbital features a substantially larger overlapping with the H s orbital, resulting in stable hydrogen adsorption structures with covalent boron-hydrogen bonding. To be more intuitive, we define a physical quantity chemical hybridization to represent the strength of chemical hybridization between B pz with H s orbitals, which is calculated by ratio of integration of H s orbitals and integration of B pz orbitals. The chemical hybridization of freestanding $\beta_{12}$ borophanes are 0.218 for $H_4$ structure, 0.147 for $H_{44}$ structure, and 0.143 for $H_{45}$ structure, respectively (Table S3). Therefore, the optimal hydrogen adsorption sites of freestanding borophenes originate from the electronic hybridization between B $p_z$ and H s orbitals. It should be mentioned that α' and $\chi_3$ borophanes precisely satisfy the similar underlying mechanism (Table S1 and S3).

To better understand the stable freestanding configurations, we investigate the borophanes with different hydrogen concentrations by varying element ratios of hydrogen and boron atoms. Compared with the lattice constant of pristine borophenes, the fully relaxed lattice constants of hydrogenated borophenes only experience a slight change (~1.4%) for a quite high element ratio, which leads to a negligible difference in electronic properties. Therefore, we employ a larger supercell using the lattice constant of pristine borophenes and consider different hydrogen densities, accompanied with the reduced element ratio of hydrogen and boron atoms (0.1, 0.125, and 0.0625 for $\beta_{12}$, $\chi_3$, and $\alpha'$ borophene, respectively).

The favored configurations of $\beta_{12}$ borophene with low element ratios are shown in Figure 3. The most energetically favorable adsorption sites are $H_4$, $H_4$, and $H_6$ of $\beta_{12}$, $\chi_3$, and $\alpha'$ borophenes, respectively, which is consistent with the medium element ratio. Note that the interactions between the two adsorbed H atoms have been taken into account in calculating $E_{ad}$ through subtraction of $E_H$, which is the total energy of the simulation cell containing two H atoms fixed at the identical sites of borophanes. Moreover, the favored borophanes are obviously dependent on the intensity of chemical hybridization between B $p_z$ and H s orbitals (Table S3 and Figure S7), regardless of the element ratios.

With the decrease of the element ratio, the metallic behavior is retained as the several bands clearly crossing the Fermi level, as shown in Figure 3d-f. Notably, for the second favored $\beta_{12}$ borophane (Figure 3e), a near linear band crossing point occurs at the Fermi level near the M point, indicating a near massless fermion. The Dirac cone-like electronic structures of borophane can be tuned through varying the element ratio of boron and hydrogen atoms. Although, the special Dirac core-like structure comes from the secondly stable structure, its energy is only 49 meV higher than that of the most favored structures, suggesting large probability of simultaneous observing these two configurations in experiments. Besides, phase transition from the most stable structure to metastable structure could be realized by light, electric and heat manipulation, in favor of exploration and utilization of this particular electronic property.

We then examine borophane configurations with higher hydrogen densities (i.e., the higher element ratios of 0.4, 0.5, and 0.25 for $\beta_{12}$, $\chi_3$, and $\alpha'$ borophenes, respectively), which correspond to configurations of two hydrogen atoms bonding on the unit cell. Due to the strong interaction between adsorbed H atoms, the H adsorption patterns change obviously with the increased element ratio, where two H atoms prefers to adsorb on opposite sides of borophenes. The optimal adsorption sites are $H_4H_5$, $H_{44}H_5$, and $H_5$ for $\beta_{12}$, $\chi_3$, and $\alpha'$ borophenes, respectively. One of the favored adsorption sites ($H_{44}$) of $\beta_{12}$ borophane is in excellent agreement with the previous work of Yu et al.[40]. Additionally, electronic structure (Figure S11) and Bader charge analysis (Table S1) indicate that the contribution of chemical hybridization takes the leading role, which accords closely with the results of low and medium element ratios.

To evaluate the thermodynamical stability of borophanes, 2 ps adiabatic molecular dynamics simulations of $\beta_{12}$ borophane at 300 K were carried out in the microcanonical ensemble (NVE) with timestep of 1 fs. Figure 4a-c exhibit snapshots of $\beta_{12}$ borophane with adsorption of $H_4$ at *t* = 0, 1 and 2 ps, implying that the hydrogen atoms stick firmly to boron sheet and the entire structure is overwhelmingly stable at room temperature. To be prudent, migration barrier for single hydrogen atoms to diffuse from $H_4$ to $H_{44}$ has been calculated using the climbing image nudged elastic band method[41]. Figure 4d shows the energy barrier of in-plane diffusion path of freestanding $β_{12}$ borophane. The high diffusion barrier (~0.43 eV) makes H atom hardly diffuses from $H_4$ to other sides at room temperature, thus enabling extremely thermodynamical stable configurations.

In recent experiments, Ag(111) is reported to be an excellent substrate to support the metallic borophene growth[42] and $β_{12}$ borophene is the most frequently observed phase. Despite such a small charge transfer between borophanes and Ag substrate, the favorable hydrogen adsorption site is modulated significantly. Interestingly, the hydrogen atoms tend to adsorb on $H_6$ and $H_4$ sites (in the sequence of relative energies) of Ag supported $β_{12}$ borophene, and the minimum hydrogen adsorption energy of -3.15 eV is reached for the $H_6$ structure (Figure 5 a and b, respectively). The obtained adsorption energies are shown in Figure S12, where the $H_6$ site is found to be the most favorable H adsorption site. Additionally, when H atoms continue to be placed on $H_4$, with the B-H bond already formed at $H_6$ sites, the strong repulsion between the two H atoms has a significant influence on the chemical hybridization between H atoms and borophene, resulting in spontaneous diffusion of hydrogen from $H_4$ to $H_{44}$ site during geometrical optimization (Figure 5c), which coincides well with previous findings[35]. In stark contrast, the recent work by Li *et al.*[32] demonstrated the lowest-energy configuration is the $H_{44}$ site, and $H_6$ is the second most favorable site by keeping the initial $H_{44}$ atom fixed on $β_{12}$ borophane, which is attributed to the structural optimization limited in the out-of-plane direction.

As further discussed in more detail in Figure S12 and Table S4, the energetically stable $\beta_{12}$ borophane structures are associated with chemical hybridization on Ag substrate, which are 0.234, 0.220, and 0.219 for $H_6$, $H_4$, and $H_5$ sites, respectively. Interestingly, Ag substrate leads to the interfacial charge transfer to boron sheets, resulting in distinct hybridization of B $p_z$ with H s orbitals and new borophane configurations. Thus, the substrate is a strong driving force towards the mediation of favored borophanes. For a broader picture, we extend our investigation to Au and Cu substrates. In the previous studies, experimental scientists investigated the fabrication mechanism of $\beta_{12}$ borophene on Au(111) substrates[43] and a new single-crystal borophene on Cu(111) surfaces[44]. Due to the specific rotation of boron sheets on substrates, large supercells were employed to match the interfacial interactions, which is beyond the scope of our work. Hence, we replace Ag substrate with Au and Cu to illustrate the optimal adsorption sites and mechanism, aiming to confirm the effect of different metal substrates. The stable borophanes on metals have similar features in H adsorption configurations and energetically favored sites ($H_6$, $H_4$, and $H_5$). From the decomposed orbital-resolved density of states, we find that the chemical hybridization contributed by B $p_z$ and H s orbitals is not affected greatly upon different metal substrates (See Figure S14 and Table S4), confirming the microscopic mechanism.

Finally, all three borophanes with the lowest $E_{ad}$ supported by Ag substrate display different $H_6$ sites adsorption for the element ratio of 0.1 (Figure S15), considering typical adsorption sites ($H_6$ and $H_4$). The chemical energy of orbital hybridization, which constructs a strong B-H bond, dominates the favorable borophanes (Figure S16). Therefore, we reveal that metal substrates can effectively mediate the favored borophane configurations and the optimal borophane configurations are not sensitively affected by the difference of metal substrates.

In summary, we investigated the energetically favored configurations of borophane with various hydrogen densities for the cases of freestanding configurations and on different metal substrates. Based on DFT calculations, the optimal $\beta_{12}$ borophanes on Ag substrates exhibit distinct configurations from freestanding borophenes. Besides, we demonstrated the energetically favored hydrogen adsorption sites under different element ratios of hydrogen and boron atoms. We revealed that it is the chemical energy of orbitals hybridization between B $p_z$ and H s orbitals that determines the lowest-energy borophane configurations. The underlying mechanism of hydrogen adsorption could be generalized on other substrates and element ratios. Moreover, this work provides an effective strategy to rationalize the identical underlying mechanism, enabling boron-based nanoelectronic materials in high-speed electronic transport and switching devices.

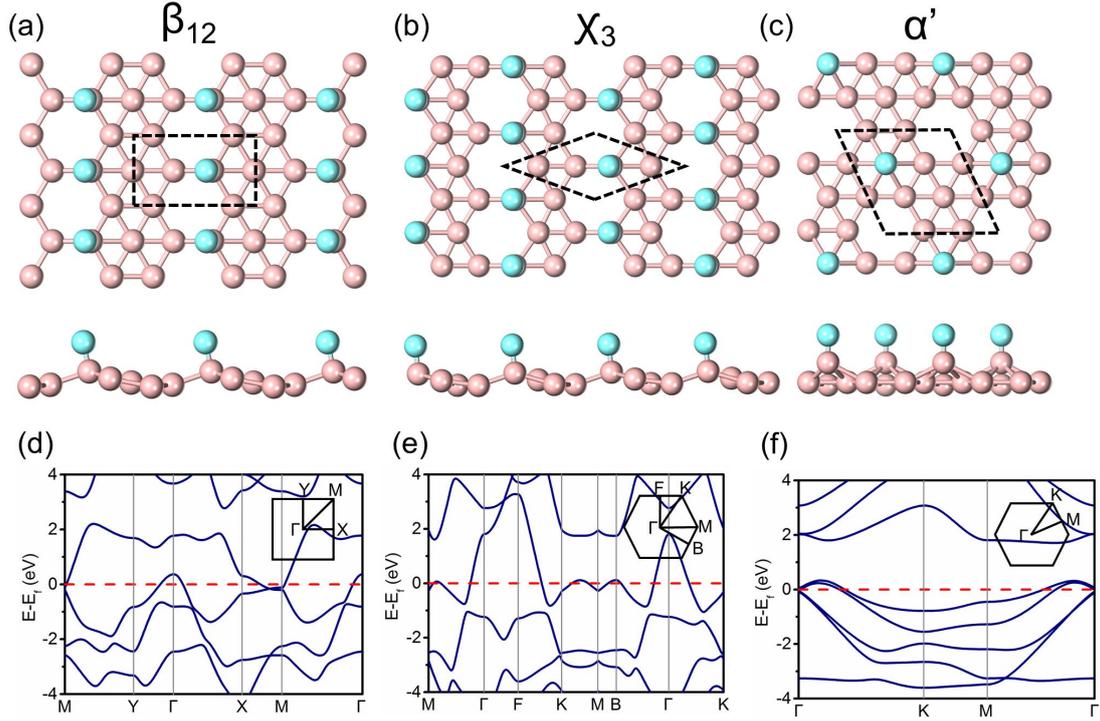

**Figure 1. The most energetically stable structures of freestanding borophanes with element ratios of 0.2, 0.25 and 0.125, respectively.** Top and side views of the most stable (a) $β_{12}$, (b) $χ_3$, and (c) $α'$ type boron monolayer sheets with hydrogen adsorption sites of $H_4$, $H_4$, and $H_6$, respectively. Calculated electronic band structures of (d) $β_{12}$ borophane, (e) $χ_3$ and (f) $α'$ borophanes, respectively. Dashed black lines in panels a-c indicate the unit cells of freestanding borophane. Boron and hydrogen atoms are distinguished with pink and cyan circles, respectively. The Brillouin zones and high-symmetry paths are denoted in panel d-f. The Fermi energy is set as zero.

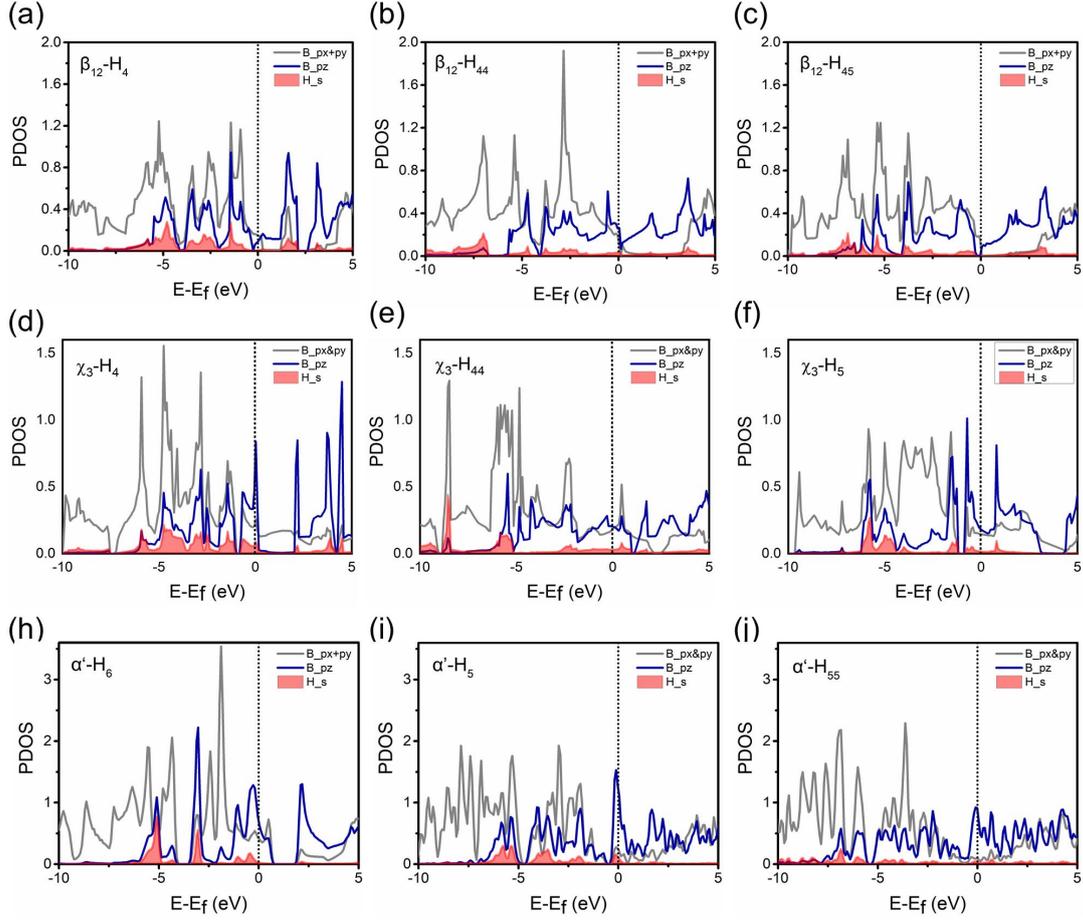

**Figure 2. Mechanism of orbitals hybridization dependence of borophane structures.** Projected density of states (PDOS) of energetically stable freestanding $\beta_{12}$ borophanes with the element ratio of 0.2, denoted as (a) $H_4$, (b) $H_{44}$, and (c) $H_{45}$, respectively. PDOS of energetically stable freestanding $\chi_3$ borophene with element ratio of 0.25, denoted as (d) $H_4$, (e) $H_{44}$, and (f) $H_5$, respectively. PDOS of energetically stable freestanding $\alpha'$ borophane with the element ratio of 0.125, denoted as d) $\alpha'$-$H_6$, e) $\alpha'$-$H_5$, and f) $\alpha'$-$H_{55}$, respectively. The dotted line at zero energy represents the Fermi level.

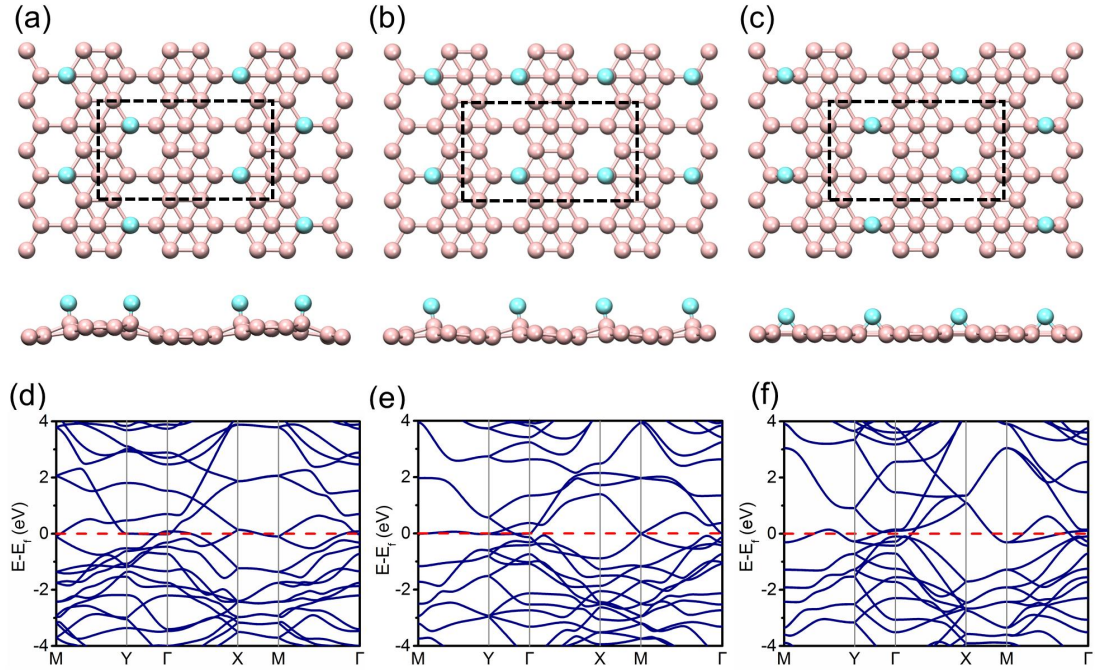

**Figure 3. Energetically stable structures of freestanding $\beta_{12}$ borophane with the element ratio of 0.1.** Top and side views of (a) the most stable freestanding $\beta_{12}$ borophane, (b) the secondly favored freestanding $\beta_{12}$ borophane and (c) the thirdly favored freestanding $\beta_{12}$ borophane. (d-f) Calculated electronic band structures of top three energetically stable freestanding $\beta_{12}$ borophane with element ratio of 0.1. The Fermi energy is set as zero in (d-f).

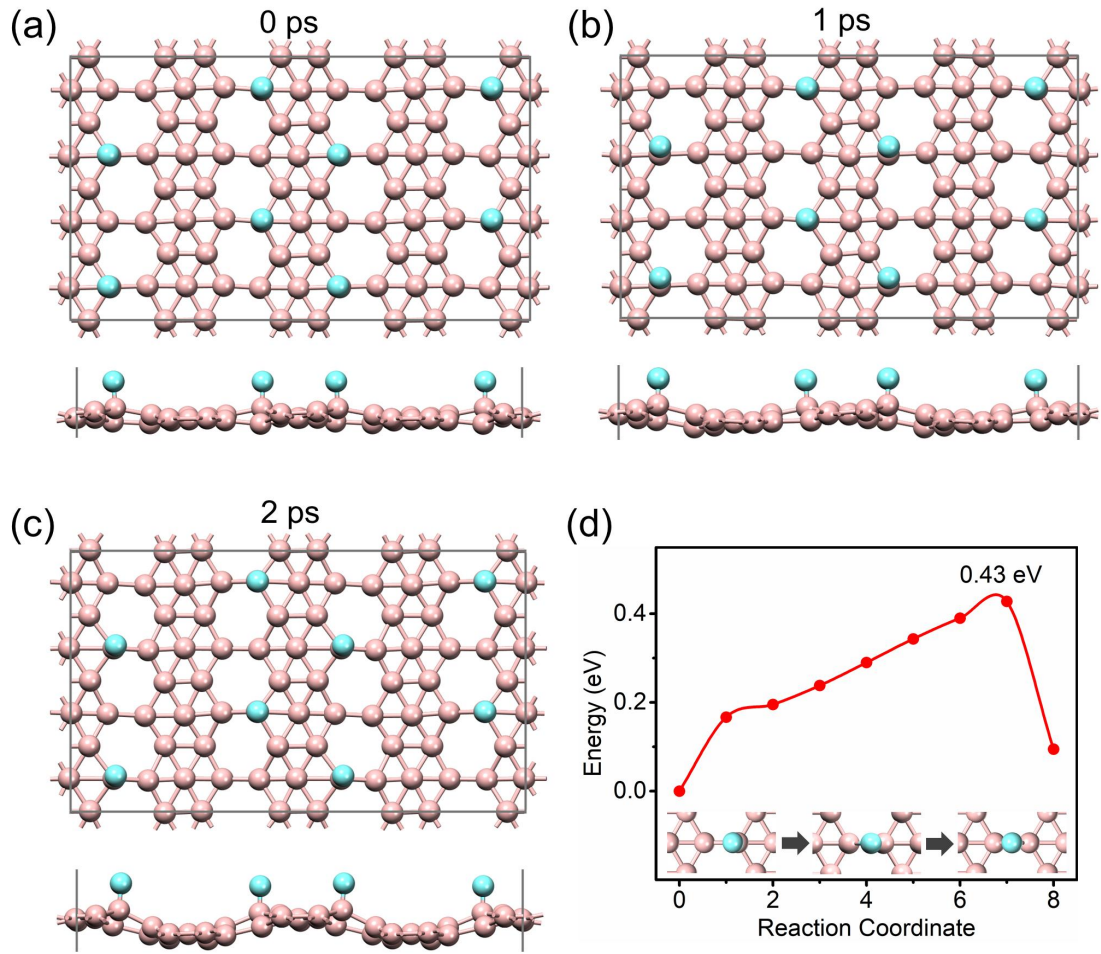

**Figure 4. Thermal dynamical stability of freestanding β$_{12}$ borophanes with element ratio of 0.1.** Snapshots of top and side views of β$_{12}$ borophane at (a) 0 ps, (b) 1 ps, and (c) 2 ps during the 2 ps NVE simulations at 300 K. (d) The energy barriers for the diffusion of one hydrogen atom from H$_4$ to H$_{44}$ sites. Insets indicate the initial, transition, and final states of freestanding β$_{12}$ borophanes.

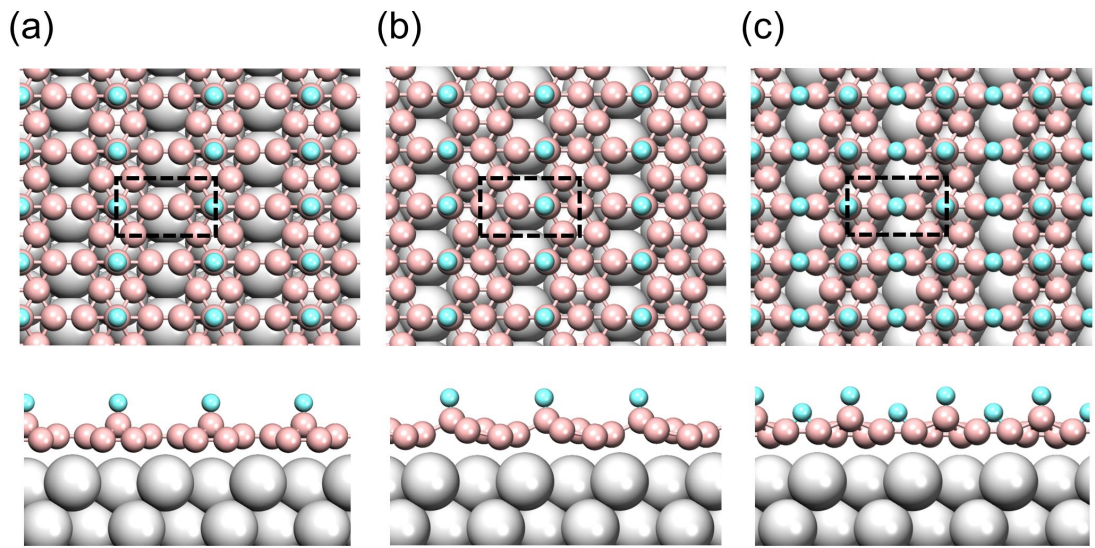

**Figure 5. Stable configurations of Ag supported β$_{12}$ borophanes.** Top and side views of the most energetically favorable β$_{12}$ borophane on Ag substrate with the element ration of 0.2, in which the H atom adsorbs on (a) H$_6$ and (b) H$_4$ sites. (c) Top and side views of the most energetically favorable β$_{12}$ borophane on Ag substrate with the element ratio of 0.4, in which the H atoms adsorb on H$_6$ and H$_{44}$ sites. The dashed rectangles indicate the unit cells of borophane on Ag substrate. Cyan, pink, and grey balls indicate hydrogen, boron, and silver atoms, respectively.

## ASSOCIATED CONTENT

**Supporting Information.**

The Supporting Information is available free of charge at

The adsorption energy of borophanes for different element ratios, Electron transfer between boron and hydrogen atoms, projected density of states, optimal metal supported borophane configurations.

## AUTHOR INFORMATION

**Corresponding Author**

Wei Ma − Ningxia Key Laboratory of Photovoltaic Materials, Ningxia University, Yinchuan 750021, People's Republic of China; Tel: + 86-0951-2062414. Email: mawei@nxu.edu.cn

Jin Zhang − Center for Free Electron Laser Science, Max Planck Institute for the Structure and Dynamics of Matter, 22761 Hamburg, Germany; orcid.org/0000-0001-7830- 3464; Email: jin.zhang@mpsd.mpg.de

Authors

Yuchong Kang − Ningxia Key Laboratory of Photovoltaic Materials, School of Materials and New Energy, Ningxia University, Yinchuan, Ningxia 750021, P.R. China

Xiaoyun Ma − Ningxia Key Laboratory of Photovoltaic Materials, School of Materials and New Energy, Ningxia University, Yinchuan, Ningxia 750021, P.R.

China

Jing Fu − Ningxia Key Laboratory of Photovoltaic Materials, School of Materials and New Energy, Ningxia University, Yinchuan, Ningxia 750021, P.R. China

Haibo Li − Ningxia Key Laboratory of Photovoltaic Materials, School of Materials and New Energy, Ningxia University, Yinchuan, Ningxia 750021, P.R. China

Kun Yang − Ningxia Key Laboratory of Photovoltaic Materials, School of Materials and New Energy, Ningxia University, Yinchuan, Ningxia 750021, P.R. China

Zongguo Wang − Ningxia Key Laboratory of Photovoltaic Materials, School of Materials and New Energy, Ningxia University, Yinchuan, Ningxia 750021, P.R. China

Author Contributions

The manuscript was written through contributions of all authors. All authors have given approval to the final version of the manuscript.

ACKNOWLEDGEMENTS

We acknowledge financial supports from the National Natural Science Foundation of China (grants 11704207, 51802312), Project of Ningxia Key R&D Plan (2018BEE03014), the Third Batch of Ningxia Youth Talents Supporting Program (TJGC2018006), J.Z. acknowledges funding received from the European Union Horizon 2020 research and innovation program under Marie Sklodowska-Curie Grant

Agreement 886291 (PeSD-NeSL).

**REFERENCES**

(1) Feng, B.; Sugino, O.; Liu, R.Y.; Zhang, J.; Yukawa, R.; Kawamura, M.; Iimori, T.; Kim, H.; Hasegawa, Y.; Li, H.; Chen, L.; Wu, K.; Kumigashira, H.; Komori, F.; Chiang, T.C.; Meng, S.; Matsud a, I. Dirac Fermions in Borophene. *Phys. Rev. Lett.* **2017**, 118, 096401.

(2) Zhou, X.-F.; Dong, X.; Oganov, A.R.; Zhu, Q.; Tian, Y.; Wang, H.-T., Semimetallic Two-Dimensional Boron Allotrope with Massless Dirac Fermions. *Phys. Rev. Lett.* **2014**, 112, 085502.

(3) Vishnoi, P.; Pramoda, K.; Rao, C.N.R. 2D Elemental Nanomaterials Beyond Graphene. *ChemNanoMat* **2019**, 5, 1062-1091.

(4) Jiang, H.R.; Lu, Z.; Wu, M.C.; Ciucci, F.; Zhao, T.S. Borophene: A promising anode material offering high specific capacity and high rate capability for lithium-ion batteries. *Nano Energy* **2016**, 23, 97-104.

(5) Zhang, J.; Zhang, J.; Zhou, L.; Cheng, C.; Lian, C.; Liu, J.; Tretiak, S.; Lischner, J.; Giustino, F.; Meng, S. Universal Scaling of Intrinsic Resistivity in Two-Dimensional Metallic Borophene. Angew. *Chem. Int. Ed.* **2018**, 57, 4585–4589.

(6) Wang, Z.-Q.; Lü, T.-Y.; Wang, H.-Q.; Feng, Y.P.; Zheng, J.-C. Band structure engineering of borophane by first principles calculations. *RSC Adv.* **2017**, 7, 47746-47752.

(7) Zhou, Y.P.; Jiang, J.W. Molecular dynamics simulations for mechanical properties of borophene: parameterization of valence force field model and Stillinger-Weber potential. *Sci. Rep.* **2017**, 7, 45516.

(8) Zhang, Z.; Yang, Y.; Penev, E. S.; Yakobson, B. I. Elasticity, Flexibility, and Ideal Strength of Borophenes. *Adv. Funct. Mater.* **2017**, 27, 16505059

(9)Penev, E. S.; Bhowmick, S.; Sadrzadeh, A.; Yakobson, B. I. Polymorphism of two-dimensional boron. *Nano Lett.* **2012**, 12, 2441-2445.

(10) Cui, H.; Zhang, X.; Chen, D. Borophene: a promising adsorbent material with strong ability and capacity for $SO_2$ adsorption. *Appl. Phys. A* **2018**, 124, 636.

(11) Rubab, A.; Baig, N.; Sher, M.; Sohail, M. Advances in ultrathin borophene materials. *Chem. Eng. J* **2020**, 401, 126109.

(12) Li, X.; Tan, X.; Xue, Q.; Smith, S. Charge-controlled switchable $H_2$ storage on conductive borophene nanosheet. *Int. J. Hydrog. Energy* **2019**, 44, 20150-20157.

(13) Qin, G.; Cui, Q.; Du, A.; Sun, Q. Borophene: A Metal-free and Metallic Electrocatalyst for Efficient Converting $CO_2$ into $CH_4$. *ChemCatChem* **2020**, 12, 1483-1490.

(14) Li, L.; Zhang, H.; Cheng, X. The high hydrogen storage capacities of Li-decorated borophene. *Comput. Mater. Sci.* **2017**, 137, 119-124.

(15) Folorunso, O.; Hamam, Y.; Sadiku, R.; Sinha Ray, S.; Adekoya, G.J. Theoretical analysis of borophene for lithium ion electrode. Mater. Today 2021, 38, 485-489.

(16) Ou, M.; Wang, X.; Yu, L.; Liu, C.; Tao, W.; Ji, X.; Mei, L. The Emergence and Evolution of Borophene. *Adv. Sci. (Weinh)* **2021**, 8, 2001801.

(17) Tatullo, M.; Zavan, B.; Genovese, F.; Codispoti, B.; Makeeva, I.; Rengo, S.; Fortunato, L.; Spagnuolo, G. Borophene Is a Promising 2D Allotropic Material for Biomedical Devices. *Appl. Sci.* **2019**, 9, 3446.

(18) Penev, E. S.; Kutana, A.; Yakobson, B. I. Can Two-Dimensional Boron Superconduct? *Nano Lett.* **2016**, 16, 2522-2526.

(19) Liu, Y., Penev, E. S., Yakobson, B. I. Probing the synthesis of two-dimensional boron by first-principles computations. *Angew. Chem. Int. Ed.* **2013**, 52, 3156–3159.

(20) Feng, B.; Zhang, J.; Liu, R.-Y.; Iimori, T.; Lian, C.; Li, H.; Chen, L.; Wu, K.; Meng, S.; Komori, F.; Matsuda, I. Direct evidence of metallic bands in a monolayer boron sheet. *Phys. Rev. B* **2016**, 94, 041408.

(21) Feng, B.; Zhang, J.; Zhong, Q.; Li, W.; Li, S.; Li, H.; Cheng, P.; Meng, S.; Chen, L.; Wu, K. Experimental realization of two-dimensional boron sheets. *Nat. Chem.* **2016**, 8, 563-568.

(22) Zhong, Q.; Zhang, J.; Cheng, P.; Feng, B.; Li, W.; Sheng, S.; Li, H.; Meng, S.; Chen, L.; Wu, K. Metastable phases of 2D boron sheets on Ag(1 1 1). *J. Phys. Condens. Matter* **2017**, 29, 095002.

(23) Vinogradov, N. A.; Lyalin, A.; Taketsugu, T.; Vinogradov, A. S.; Preobrajenski, A. Single-Phase Borophene on Ir(111): Formation, Structure, and Decoupling from the Support. *ACS Nano* **2019**, 13, 14511.

(24) Wu, R.; Drozdov, I.K.; Eltinge, S.; Zahl, P.; Ismail-Beigi, S.; Bozovic, I.; Gozar, A. Large-area single-crystal sheets of borophene on Cu(111) surfaces. *Nat. Nanotechnol* **2019**, 14, 44-49.

(25) Zhang, Z.; Yang, Y.; Gao, G.; Yakobson, B.I. Two-Dimensional Boron Monolayers Mediated by Metal Substrates. *Angew. Chem. Int. Ed. Engl* **2015**, 54, 13022-13026.

(26) Zhang, Z.; Penev, E. S.; Yakobson, B. I. Two-dimensional boron: Structures, properties and applications. *Chem. Soc. Rev.* **2017**, 46, 6746–6763.

(27) Peng, B.; Zhang, H.; Shao, H.; Ning, Z.; Xu, Y.; Ni, G.; Lu, H.; Zhang, D.W.; Zhu, H. Stability and strength of atomically thin borophene from first principles calculations. *Mater. Res. Lett.* **2017**, 5, 399-407.

(28) Mortazavi, B.; Dianat, A.; Rahaman, O.; Cuniberti, G.; Rabczuk, T. Borophene as an anode material for Ca, Mg, Na or Li ion storage: A first-principle study. *J. Power Sources* **2016**, 329, 456-461.

(29) Hou, C.; Tai, G.; Hao, J.; Sheng, L.; Liu, B.; Wu, Z. Ultrastable Crystalline Semiconducting Hydrogenated Borophene. *Angew. Chem. Int. Ed. Engl* **2020**, 59, 10819-10825.

(30) Lei, X.; Zatsepin, A.F.; Boukhvalov, D.W. Chemical instability of free-standing boron monolayers and properties of oxidized borophene sheets. *Physica E Low Dimens* **2020**, 120, 114082.

(31) Balog, R.; Jorgensen, B.; Nilsson, L.; Andersen, M.; Rienks, E.; Bianchi, M.; Fanetti, M.; Laegsgaard, E.; Baraldi, A.; Lizzit, S.; Sljivancanin, Z.; Besenbacher, F.; Hammer, B.; Pedersen, T.G.; Hofmann, P.; Hornekaer, L. Bandgap opening in graphene induced by patterned hydrogen adsorption. *Nat. Mater.* **2010**, 9, 315-9.

(32) Li, Q.; Kolluru, V. S. C.; Rahn, M. S.; Schwenker, E.; Li, S.; Hennig, R. G.; Darancet, P.; Chan, M. K. Y.; Hersam, M. C. Synthesis of borophane polymorphs through hydrogenation of borophene. *Science* **2021**, 371, 1143–1148.

(33) Pang, K.; Xu, X.; Ku, R.; Wei, Y.; Ying, T.; Li, W.; Yang, J.; Li, X.; Jiang, Y.

First-Principles Calculations for the Impact of Hydrogenation on the Electron Behavior and Stability of Borophene Nanosheets: Implications for Boron 2D Electronics. *ACS Appl. Nano Mater.* **2022**, 5, 1419-1425.

(34) Tai, G.; Xu, M.; Hou, C.; Liu, R.; Liang, X.; Wu, Z., Borophene Nanosheets as High-Efficiency Catalysts for the Hydrogen Evolution Reaction. *ACS Appl. Mater. Interfaces* **2021**, 13, 60987-60994.

(35) Xu, Y.; Zhang, P.; Xuan, X.; Xue, M.; Zhang, Z.; Guo, W.; Yakobson, B.I. Borophane Polymorphs. *J. Phys. Chem. Lett.* **2022**, 13, 1107-1113.

(36) Tang, H.; Ismail-Beigi, S. Novel precursors for boron nanotubes: the competition of two-center and three-center bonding in boron sheets. *Phys. Rev. Lett.* **2007**, 99, 115501.

(37) Wu, X.; Dai, J.; Zhao, Y.; Zhuo, Z.; Yang, J.; Zeng, X. C. Two-dimensional boron monolayer sheets. *ACS Nano* **2012**, 6, 7443–7453.

(38) Princy Maria, J.; Bhuvaneswari, R.; Nagarajan, V.; Chandiramouli, R. Surface adsorption studies of benzyl bromide and bromobenzyl cyanide vapours on black phosphorene nanosheets – a first-principles perception. *Mol. Phys.* **2020**, 118, 1737744.

(39) Haque, E.; Stampfl, C.; Hossain, M. A. Prediction of the fundamental properties of novel Be-B-Ta-based ternary compounds from first-principles calculations. *Phys. Rev. Mater.* **2019**, 3, 084804.

(40) Yu, M.; Zhang, Z.; Guo, W., Structures, Mechanics, and Electronics of Borophanes. *J. Phys. Chem. C* **2021**, 125, 22917-22928.

(41) Henkelman, G.; Uberuaga, B. P.; Jónsson, H. Climbing Image Nudged Elastic Band Method for Finding Saddle Points and Minimum Energy Paths. *J. Chem. Phys.* **2000**, 113, 9901–9904.

(42) Mannix, A. J., Zhou, X. F., Kiraly, B., Wood, J. D., Alducin, D., Myers, B. D., Liu, X., Fisher, B. L., Santiago, U., Guest, J. R., Yacaman, M. J., Ponce, A., Oganov, A. R., Hersam, M. C., Guisinger, N. P. Synthesis of borophenes: Anisotropic, two-dimensional boron polymorphs. *Science* **2015**, 350, 1513–1516.

(43) Kiraly, B., Liu, X., Wang, L., Zhang, Z., Mannix, A. J., Fisher, B. L., Yakobson, B. I., Hersam, M. C., Guisinger, N. P. Borophene Synthesis on Au(111). *ACS Nano* **2019**, 13, 3816–3822.

(44) Wu, R., Drozdov, I. K., Eltinge, S., Zahl, P., Ismail-Beigi, S., Božović, I., Gozar, A. Large-area single-crystal sheets of borophene on Cu(111) surfaces. *Nat. Nanotechnol.* **2019**, 14, 44–49.

**TOC Figure**

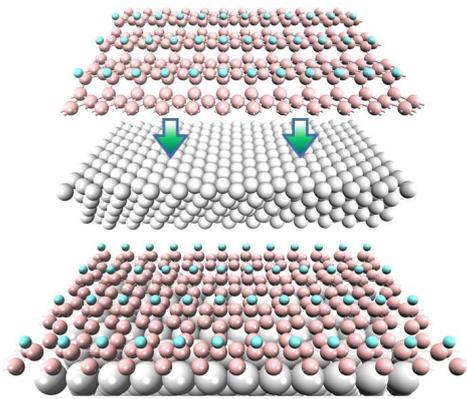